\documentclass[onecolumn, showpacs]{revtex4}

\usepackage{graphicx}
\usepackage{dcolumn}
\usepackage{bm}

\input epsf

\begin{document}

\title{\Large A  Quintessence  Problem  in  Brans-Dicke Theory \\ with
Varying  Speed  of  Light }

\author{\bf Subenoy Chakraborty}
\email{subenoyc@yahoo.co.in}
\author{\bf N. C. Chakraborty  }
\author{\bf Ujjal Debnath}
\email{ujjaldebnath@yahoo.com}

\affiliation{Department of Mathematics, Jadavpur University,
Calcutta-32, India.}

\date{\today}

\begin{abstract}
It  is  shown  that  minimally  coupled  scalar  field  in
Brans-Dicke  theory  with  varying  speed  of  light  can  solve
the  quintessence  problem  and  it  is  possible  to  have  a
non-decelerated  expansion of  the  present  universe  with
BD-theory  for  anisotropic  models  without  any  matter.
\end{abstract}

\maketitle

\section{\normalsize\bf{Introduction}}
Recently, the  matter  field  can  give  rise  to  an
accelerated  expansion [1]  for  the  universe stems  from  the
observational  data  regarding  the  luminosity-redshift
relation  of  type Ia  supernovae [2,3]  up  to  about $z\sim 1$.
This  matter  field  is  called  'quintessence matter' (shortly
Q-matter). This  Q-matter  can  behave  like a cosmological
constant [4,5]  by  combining  positive  energy density  and
negative  pressure. So there  must  be  this Q-matter  either
neglected  or  unknown  responsible  for  this accelerated
universe. At  the  present  epoch, a  lot  of works has been
done  to  solve  this  quintessence  problem and most popular
candidates  for  Q-matter  has  so  far  been a scalar field
having  a  potential  which  generates  a sufficient negative
pressure. Since  in  a  variety  of inflationary models scalar
fields  have  been  used  in describing  the transition from the
quasi-exponential expansion  of  the  early universe to  a
power-law  expansion, it  is  natural  to  try to understand the
present acceleration  of  the  universe, which has  an
exponential behaviour  too, by  constructing models where  the
matter responsible  for  such  behaviour  is also represented by
a scalar  field [6]. However, now  we deal with the opposite
task, i.e., we  would  like  to describe the transition from a
universe  filled  with  dust-like matter  to an exponentially
expansion  universe  and  scalar fields  are not the  only
possibility  but  there  are  (of course) alternatives.
Different  forms  for  the  quintessence energy have  been
proposed. They  include  a  cosmological constant (or,  more
generally  a  variable  cosmological  term) a scalar field  [4,7]
a  frustrated  network  of  non-abelian cosmic strings  and  a
frustrated  network  of  domain  wall [8,9]. All  these proposal
assume  the  Q-matter  behaves  as a perfect  fluid with  a
linear  barotropic  equation  of state and  so  some effort  has
been  invested  in determining its adiabatic  index at  the
present  epoch  [10,11]. Recently, Chimento  et al [12] showed
that  a  combination of dissipative  effects  such as  a bulk
viscous  stress and  a quintessence  scalar  field gives an
accelerated expansion for an  open  universe  ( $k = -1$). Very
recently, Banerjee et al [13]  also  have  shown that it is
possible to  have an accelerated  universe  with BD-theory in
Friedmann  model without  any  matter.\\

The  possibility  that  the  speed  of  light  $c$  might  vary
has  recently  attracted  considerable  attention  [14-20]. In a
cosmological  setting, the  variations  in  $c$  have been shown
to  solve  the  cosmological  puzzles - the  horizon, flatness
and  Lambda  problems  of  big-bang cosmology. The variations of
velocity  of  light  can  also  solve  the quasi-lambda problems.
For  power-law  variations  in  the velocity  of light  with the
cosmological  scale  factors, Barrow  et al [16]  have shown
that   flatness   problem can be  solved. The  Machian VSL
scenario  in   which $c=c_{0}a^{n}$, introduced  by  Barrow [15]
has  significant advantages  to the   phase   transition
scenario  in   which the  speed  of light  changes  suddenly
from  $c^{-}$  to $c^{+}$, preferred by  Albrecht and  Magueijo
[14]. For changing $c$, the geometry of  the universe  is  not
affected . We  have allowed  a changing  $c$ to  do  the  job
normally done  by ``superluminal expansion". The  basic
assumpssion  is that  a variable  $c$ does  not induce
corrections  to  curvature in the cosmological frame and  that
Einstein's  equations, relating curvature  to stress  energy are
still  valid. The rationale behind  this postulate  is that  $c$
changes  in  the local Lorentzian frames  associated with
cosmological expansion. The effect is a  special relativistic
effect, not a gravitational effect. Therefore curvature  should
not feel  a changing $c$. In  a cosmological setting  the
postulate proposed implies that Brans-Dicke  field equations
remains valid even when $\dot{c}\neq 0$. Magueijo  et al [21]
find  the metrices  and variations in $c$ associated with  the
counterpart   of black holes, the outside of  star  and steller
collapse . The variation  upon the  theme are  VSL theories which
explicitely break  local Lorentz invariance, such  as  the one
proposed by Albrecht and Magueijo [14] and for  which black hole
solutions
remain  exclusive.\\

This  paper  investigates  the  possibility  of  obtaining  a
non-decelerating  expansion  ($q\leq 0$)  for  the  universe  in
BD theory  with  varying  speed  of  light  in  anisotropic
models of  the  universe.

\section{\normalsize\bf{Field  equations  and  solutions }}

We  consider  the  line-element  of  anisotropic  space-time model

\begin{equation}
ds^{2}=-c^{2}dt^{2}+a^{2}dx^{2}+b^{2}d\Omega_{k}^{2}
\end{equation}

where  $a , b$ are  functions  of  time  only  and

\begin{eqnarray}d\Omega_{k}^{2}= \left\{\begin{array}{lll}
dy^{2}+dz^{2}, ~~~~~~~~~~~~ \text{when} ~~~k=0 ~~~~~( \text{Bianchi ~I ~model})\\
d\theta^{2}+sin^{2}\theta d\phi^{2}, ~~~~~ \text{when} ~~~k=+1~~
( \text{Kantowaski-Sachs~ model})\\
d\theta^{2}+sinh^{2}\theta d\phi^{2}, ~~~ \text{when} ~~~k=-1 ~~(
\text{Bianchi~ III~ model})\nonumber
\end{array}\right.
\end{eqnarray}

Here  $k$  is  the spatial  curvature  index, so  that  the
above  three  types [22] models are Euclidean, closed  and
semi-closed  respectively.\\

Now, the  BD-field  equations  with  varying  speed  of  light are

\begin{equation}
\frac{\ddot{a}}{a}+2\frac{\ddot{b}}{b}=-\frac{8\pi}{(3+2\omega)\phi}\left[
(2+\omega)\rho_{_{f}}+3(1+\omega)\frac{p_{_{f}}}{c^{2}}\right]-\omega\left(\frac{\dot{\phi}}{\phi}
\right)^{2}-\frac{\ddot{\phi}}{\phi}
\end{equation}

\begin{equation}
\left(\frac{\dot{b}}{b}
\right)^{2}+2\frac{\dot{a}}{a}\frac{\dot{b}}{b}=\frac{8\pi\rho_{_{f}}}{\phi}-\frac{kc^{2}}{b^{2}}-\left(\frac{\dot{a}}{a}+2\frac{\dot{b}}{b}
\right)\frac{\dot{\phi}}{\phi}+\frac{\omega}{2}\left(\frac{\dot{\phi}}{\phi}
\right)^{2}
\end{equation}

and  the  wave  equation  for  the  BD  scalar  field  $\phi$  is

\begin{equation}
\ddot{\phi}+\left(\frac{\dot{a}}{a}+2\frac{\dot{b}}{b}
\right)\dot{\phi}=\frac{8\pi}{3+2\omega}\left(\rho_{_{f}}-\frac{3p_{_{f}}}{c^{2}}
\right)
\end{equation}

Here  the  velocity  of  light  $c$  is  an  arbitrary  function
of  time, $\omega$  is  the  BD  coupling  parameter.
$\rho_{_{f}}$ and $p_{_{f}}$ are density  and  hydrostatic
pressure respectively  of the fluid distribution  with  barotropic
equation  of  state

\begin{center}
$p_{_{f}}=(\gamma_{_{f}}-1)\rho_{_{f}}$
\end{center}

( $\gamma_{_{f}}$   being  the  constant  adiabatic  index  of
the  fluid with  $0\leq \gamma_{_{f}}\leq 2$ ).\\

From  the  above  field  equations, we  have  the
`non-conservation'  equation

\begin{equation}
\dot{\rho_{_{f}}}+\left(\frac{\dot{a}}{a}+2\frac{\dot{b}}{b}
\right)\left(\rho_{_{f}}+\frac{p_{_{f}}}{c^{2}}
\right)=\frac{kc\dot{c}}{4\pi b^{2}}\phi
\end{equation}

As  at  the present  epoch  the  universe  is  filled  with cold
matter (dust)  with  negligible pressure, so  using
$p_{_{f}}=0$, the equation  (4)  and  (5)  gives

\begin{equation}
\frac{d}{dt}(ab^{2}\dot{\phi})=\frac{8\pi}{3+2\omega}ab^{2}\rho_{_{f}}
\end{equation}

and

\begin{equation}
\frac{d}{dt}(ab^{2}\rho_{_{f}})=\frac{ka\phi}{8\pi}\frac{d}{dt}(c^{2})
\end{equation}

Now  eliminating $\rho_{_{f}}$  between  (6)  and  (7), we  have

\begin{equation}
\frac{d^{2}}{dt^{2}}(ab^{2}\dot{\phi})=\frac{k}{3+2\omega}a\phi\frac{d}{dt}(c^{2})
\end{equation}

Since  the  scale  factors, scalar  field  and  velocity  of
light  are  time  dependent, so we  shall  consider  the
following  cases  to  obtain  exact  analytic  form  for  the
variables  assuming  some  of  them  in  polynomial  form  or in
exponential  form.\\

{\it Case I} :  The  power  law  form  of  scale  factors  $a , b$
and BD  scalar  field  $\phi $  are  assumed  as

\begin{equation}
a(t)=a_{_{0}}t^{\alpha}, b(t)=b_{_{0}}t^{\beta},
\phi(t)=\phi_{_{0}}t^{\mu}
\end{equation}

where $a_{_{0}}, b_{_{0}},\phi_{_{0}}$ are   positive   constants
and  $\alpha, \beta, \mu$ are real   constants  with   the
restriction $\alpha+2\beta\geq 3$ (for accelerating  universe).\\

If  we  use  the  relations  (9)  in  (8), we  have

\begin{equation}
c=c_{_{0}}t^{\beta-1}
\end{equation}

where  the  positive  constant  $c_{_{0}}$  is  restricted as

\begin{equation}
b_{_{0}}^{2}\mu(\alpha+2\beta+\mu-1)(\alpha+2\beta+\mu-2)=\frac{2kc_{_{0}}^{2}(\beta-1)}{3+2\omega}
\end{equation}

Also  from  (6)  the  expression  for  $\rho_{_{f}}$ is

\begin{equation}
\rho_{_{f}}=\rho_{_{0}}t^{\mu-2}
\end{equation}

with
$$
\rho_{_{0}}=\frac{3+2\omega}{8\pi}\phi_{_{0}}\mu(\alpha+2\beta+\mu-1)
$$

For  consistency  of  the  field  equations  the  restriction
between  the  parameters  is

\begin{equation}
\left(\alpha+2\beta+\mu-\frac{1}{2}\right)^{2}=\frac{\mu(\alpha+2\beta+\mu-1)}{\beta-1}[(2+\omega)
(1-\beta)-(\alpha-\mu)(3+2\omega)]+\frac{1}{4}
\end{equation}

{\it Case II} :  The  exponential  form  of  scale  factors  and
BD scalar  field  are  assumed  as

\begin{equation}
a(t)=a_{_{0}}e^{\alpha t}, b(t)=b_{_{0}}e^{\beta t},
\phi(t)=\phi_{_{0}}e^{\mu t}
\end{equation}

where  $a_{_{0}}, b_{_{0}},\phi_{_{0}}$  are  positive constants
and $\alpha, \beta, \mu$  are  real  constants.\\

From  (8), we have
\begin{equation}
c=c_{_{0}}e^{\beta t}
\end{equation}

where  the  positive  constant  $c_{_{0}}$   will  satisfy

\begin{equation}
(3+2\omega)\mu(\alpha+2\beta+\mu)^{2}=\frac{4k\beta
c_{_{0}}^{2}}{b_{_{0}}^{2}}
\end{equation}

From  (6), the  expression  for  $\rho_{_{f}}$  is
\begin{equation}
\rho_{_{f}}=\rho_{_{0}}e^{\mu t}
\end{equation}

with~~~$\rho_{_{0}}=\frac{(3+2\omega)}{8\pi}\phi_{_{0}}\mu(\alpha+2\beta+\mu)$.\\

From  field  equations  the  relation  between  the  parameters
becomes

\begin{equation} \text{either}~~~~ \alpha+2\beta+\mu=0
\end{equation}
\begin{equation}
\text{or}~~~~(\alpha+2\beta+\mu)\left\{1+\frac{(3+2\omega)\mu}{\beta^{2}}\right\}=\mu(4+3\omega)
\end{equation}

In  both  the  cases  we  have  the  deceleration  parameter

\begin{center}
$q=-1\le 0$
\end{center}

\begin{figure}
\includegraphics[height= 1.4in]{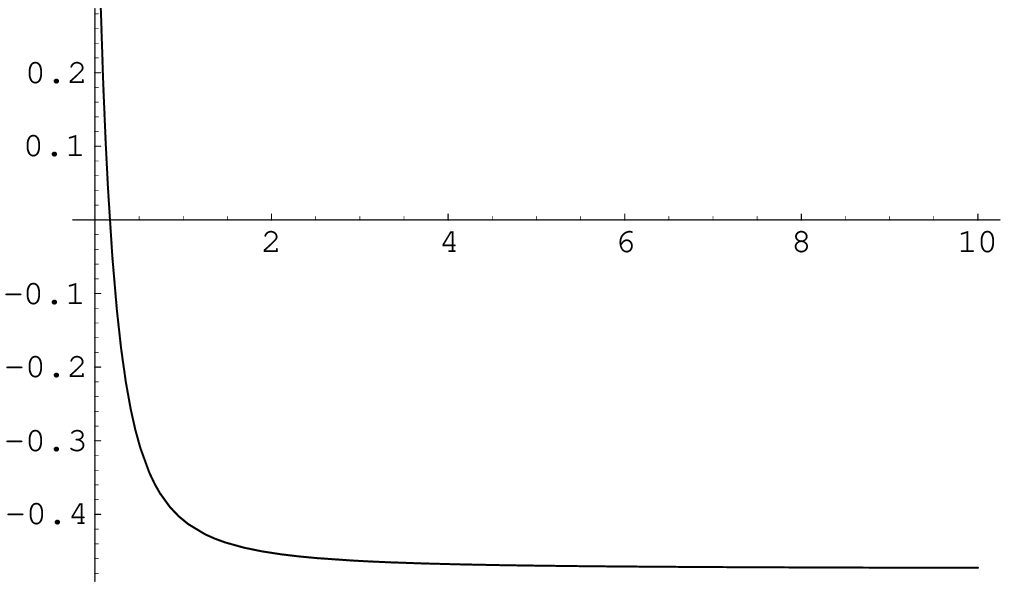}~~~~
\includegraphics[height= 1.4in]{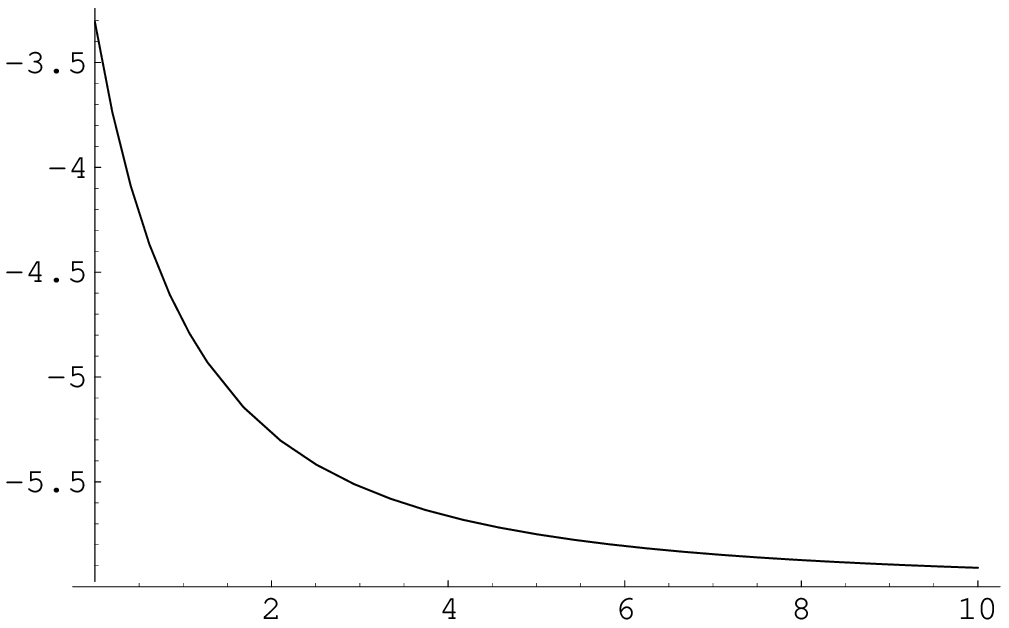}
\\\vspace{5mm}
Fig.1~~~~~~~~~~~~~~~~~~~~~~~~~~~~~~~~~~~~~~~~~~~~~~Fig.2\\
\vspace{10mm}
\includegraphics[height= 1.4in]{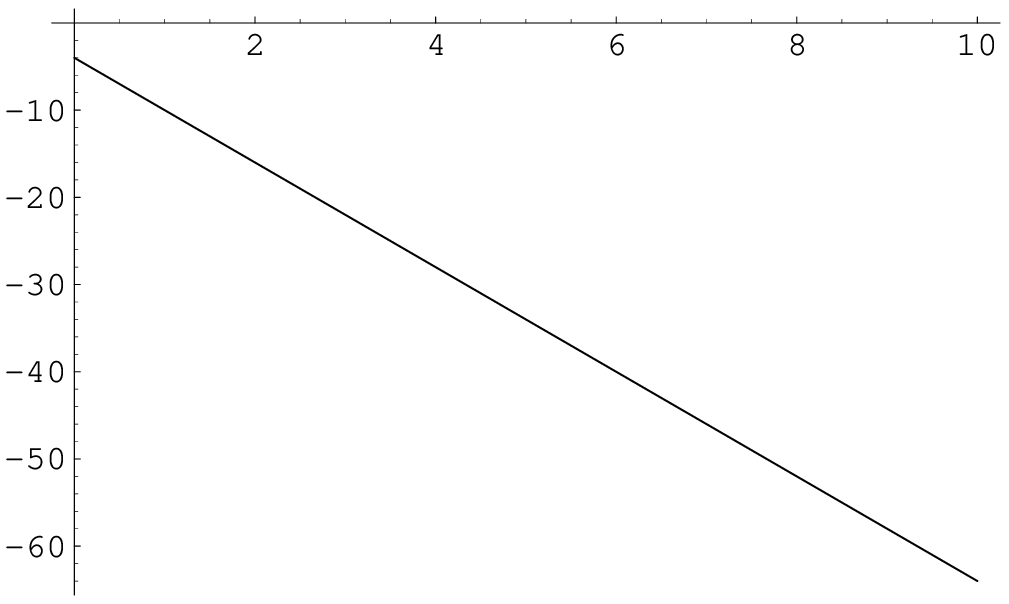}~~~~
\includegraphics[height= 1.4in]{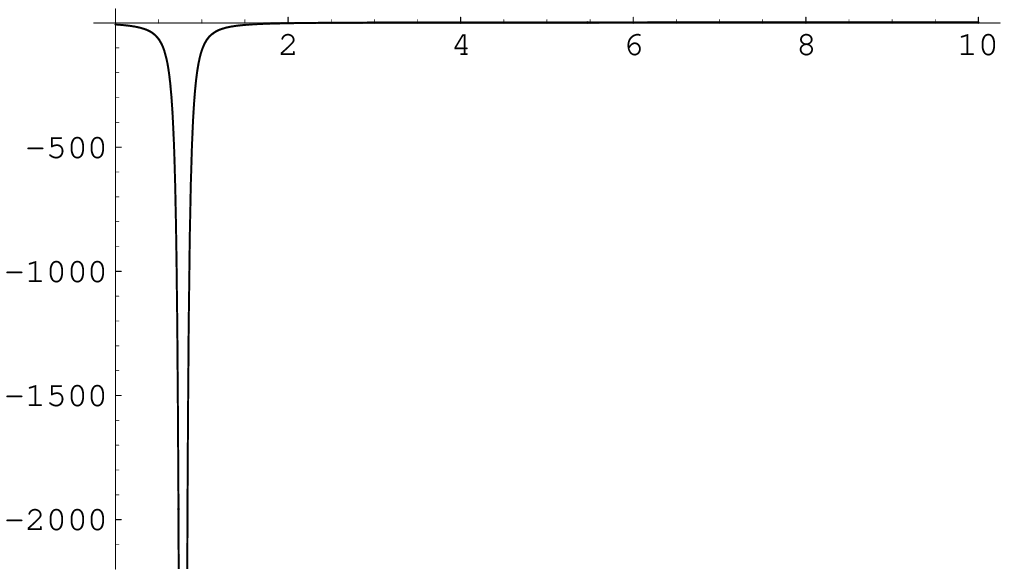}
\\\vspace{5mm}
Fig.3~~~~~~~~~~~~~~~~~~~~~~~~~~~~~~~~~~~~~~~~~~~~~~Fig.4\\
\vspace{10mm}
\includegraphics[height= 1.4in]{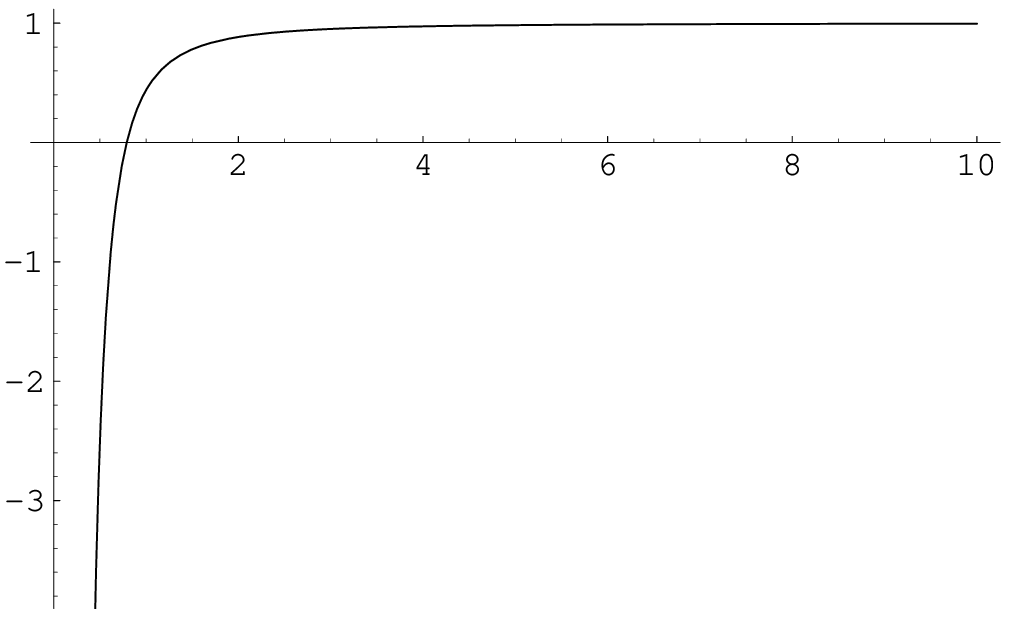}~~~~
\includegraphics[height= 1.8in]{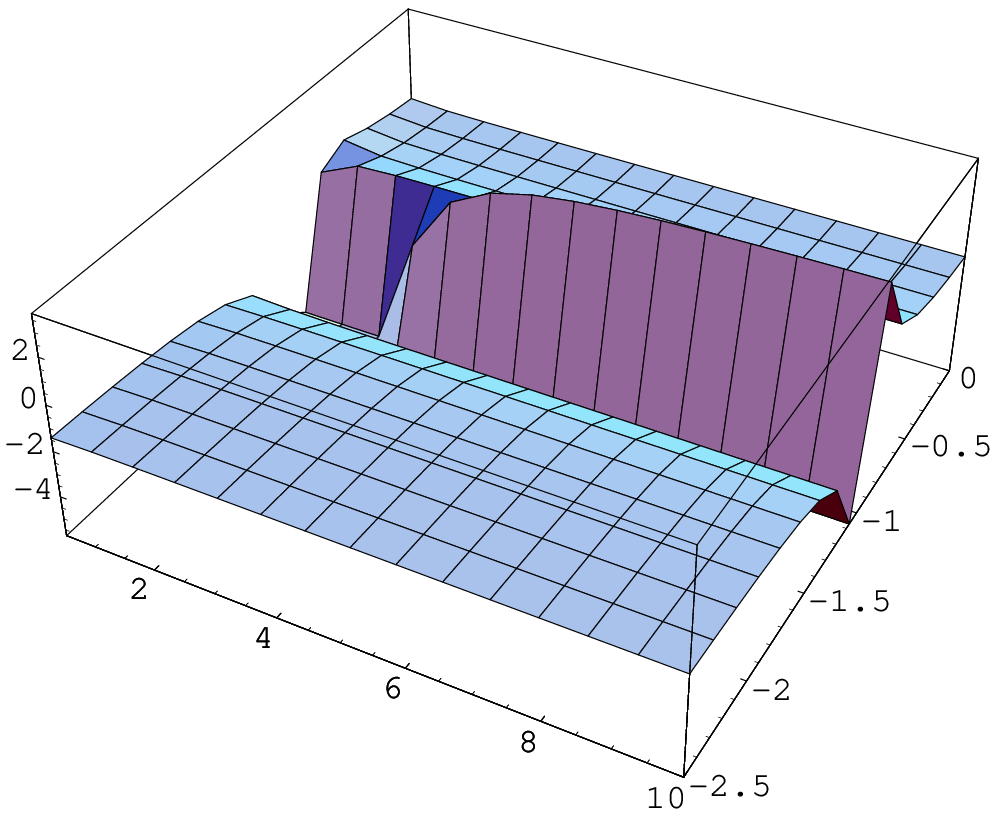}
\\\vspace{5mm}
Fig.5~~~~~~~~~~~~~~~~~~~~~~~~~~~~~~~~~~~~~~~~~~~~~~Fig.6\\
\vspace{5mm}

In Figs.1-5, we have shown the variations of $q$ over $t$ for
various values of $\omega$ and the values of parameters
$a_{0}=a_{1}=1$. We have taken $\omega=-1.8, -1.5, -1, -0.87,
-0.75 $ respectively in Figs.1-5. In Fig.6, we have shown the
variations of $q$ over $t$ and $\omega$ in the range
$0\le t\le 10$and $-2.5\le \omega \le 0$.\\
\end{figure}

{\it Case III} :   In  this  case , we  have  assumed  the  power
law form  of  scale  factor  $b$, velocity  of  light  $c$  and
BD scalar  field $\phi$  to be

\begin{equation}
b(t)=b_{_{0}}t^{\beta},c(t)=c_{_{0}}t^{\delta},
\phi(t)=\phi_{_{0}}t^{\mu}
\end{equation}

Using  (20), we  have  from  (8),

\begin{equation}
\frac{d^{2}}{dt^{2}}\left(a t^{2\beta+\mu-1}
\right)=\frac{2\delta k c_{_{0}}^{2}}{(3+2\omega)\mu
b_{_{0}}^{2}}\left(a t^{\mu+2\delta-1} \right)
\end{equation}

To  solve  the  differential  equation  we  assume
$\delta=\beta$. It is to  be  noted  that  the  above
differential  equation  can be solved  for  all  $k$. But  for
$k=\pm 1$ the  solutions  are  not consistent  to  the  field
equations. So  we  consider  only  $k = 0$. In  this  case  the
explicit solution  is

\begin{eqnarray}
\left.\begin{array}{llll}
a(t)=a_{_{0}}t^{-\frac{\omega+2}{3\omega+4}}+a_{_{1}}t^{\frac{2(\omega+1)}{3\omega+4}}\\\\
b(t)=b_{_{0}}t^{\frac{2(\omega+1)}{3\omega+4}}\\\\
c(t)=c_{_{0}}t^{\frac{2(\omega+1)}{3\omega+4}}\\\\
\phi(t)=\phi_{_{0}}t^{\frac{2}{3\omega+4}}\\\\
\rho_{_{f}}=\frac{1}{4\pi}\frac{2\omega+3}{3\omega+4}\frac{\phi_{_{0}}a_{_{1}}}{(a_{_{0}}
+a_{_{1}}t)}~t^{-\frac{3\omega+2}{3\omega+4}}
\end{array}\right\}
\end{eqnarray}

with $a_{_{0}}, a_{_{1}}$   as  integration  constants.\\

\newpage

The  deceleration  parameter  has  the  expression

\begin{center}
$q=-1+\frac{3[(3\omega+2)(a_{_{0}}+a_{_{1}}t)^{2}+a_{_{1}}^{2}(3\omega+4)t^{2}]}
{[(3\omega+2)(a_{_{0}}+a_{_{1}}t)+a_{_{1}}(3\omega+4)t]^{2}}$ ,
\end{center}

which  is  finite  for  all  $t$  and $\omega$  except
$t=-\frac{a_{_{0}}(3\omega+2)}{6a_{_{1}}(\omega+1)}$ and we have
the non-decelerated universe for $-2\le \omega\le-2/3$. Also  we
have the non-decelerated  universe asymptotically except  for
$-1\le\omega\le-1/2$. The variation  of   $q$  over  time has
been shown graphically for different  values  of  $\omega$   in
figures 1 - 5.

\section{\normalsize\bf{Conformal  transformation : Flatness  problem }}

One important  aspect  of  this model  is  that  it  can solve the
flatness  problem  as  well. To  see  this  we  make  a
conformal  transformation [23]  as

\begin{equation}
\bar{g}_{\mu\nu}=\phi g_{\mu\nu}
\end{equation}

which  enables  us  to  identify  the  energy  contributions
from  different  components  of  matter  very  clearly.\\

In  this  section, we  have  developed  the  BD  theory  in
Jordan  frame  and  to  introduce  the  Einstein  frame, we make
the  following  transformations :

\begin{equation}
d\eta=\sqrt{\phi}~a, \bar{a}=\sqrt{\phi}~a, \bar{b}=\sqrt{\phi}~b,
\psi=\text{ln}~\phi,
\bar{\rho}_{_{f}}=\phi^{-2}\rho_{_{f}},\bar{\rho}_{_{\psi}}=\phi^{-2}\rho_{_{\psi}},
\bar{p}_{_{f}}=\phi^{-2}p_{_{f}},\bar{p}_{_{\psi}}=\phi^{-2}p_{_{\psi}}
\end{equation}

So  the  field  equations  (2) - (4)  transformed  to

\begin{equation}
\frac{\bar{a}''}{\bar{a}}+2\frac{\bar{b}''}{\bar{b}}=-4\pi\left(\bar{\rho}_{_{f}}+\frac{3\bar{p}_{_{f}}}{c^{2}}\right)-\frac{(3+2\omega)}{2}\psi'^{2}
\end{equation}

\begin{equation}
\left(\frac{\bar{b}'}{\bar{b}}\right)^{2}+2\frac{\bar{a}'}{\bar{a}}\frac{\bar{b}'}{\bar{b}}+\frac{k
c^{2}}{\bar{b}^{^{2}}}=8\pi\bar{\rho}_{_{f}}+\frac{(3+2\omega)}{4}\psi'^{2}
\end{equation}

and

\begin{equation}
\psi''+\left(\frac{\bar{a}'}{\bar{a}}+2\frac{\bar{b}'}{\bar{b}}\right)\psi'=\frac{8\pi}
{3+2\omega}\left(\bar{\rho}_{_{f}}-\frac{3\bar{p}_{_{f}}}{c^{2}}\right)
\end{equation}

where~~  $' \equiv \frac{d}{d\eta}$.\\

The  scalar  field  $\psi$ (massless)  behaves  like  a  `stiff'
perfect  fluid  with  equation  of  state

\begin{equation}
\bar{p}_{_{\psi}}=\bar{\rho}_{_{\psi}}=\frac{\psi'^{2}}{16\pi G}
\end{equation}

If  the  velocity  of  light  is  constant, then  in  Einstein
frame  total  stress-energy  tensor  is  conserved  but  there
is  an  exchange  of  energy  between  the  scalar  field  and
normal  matter  according  to  the  following  equation

\begin{equation}
\bar{\rho}_{_{f}}+\left(\frac{\bar{a}'}{\bar{a}}+2\frac{\bar{b}'}{\bar{b}}\right)
\left(\bar{\rho}_{_{f}}+\frac{\bar{p}_{_{f}}}{c^{2}}\right)=-\left[\bar{\rho}_{_{\psi}}+
\left(\frac{\bar{a}'}{\bar{a}}+2\frac{\bar{b}'}{\bar{b}}\right)\left(\bar{\rho}_{_{\psi}}+
\bar{p}_{_{\psi}}\right)\right]=-\frac{\psi'}{2}\left(\bar{\rho}_{_{f}}-
\frac{3\bar{p}_{_{f}}}{c^{2}}\right)
\end{equation}

On  the  other  hand, if  the  velocity  of  light  varies then
we  have  separate  `non-conservation'  equations

\begin{equation}
\bar{\rho}_{_{f}}+\left(\frac{\bar{a}'}{\bar{a}}+2\frac{\bar{b}'}{\bar{b}}\right)\left(\bar{\rho}_{_{f}}+\frac{\bar{p}_{_{f}}}{c^{2}}\right)=-\frac{\psi'}{2}\left(\bar{\rho}_{_{f}}-\frac{3\bar{p}_{_{f}}}{c^{2}}\right)+\frac{k
c c'}{4\pi G\bar{b}^{2}}
\end{equation}

and

\begin{equation}
\bar{\rho}_{_{\psi}}+\left(\frac{\bar{a}'}{\bar{a}}+2\frac{\bar{b}'}{\bar{b}}\right)\left(\bar{\rho}_{_{\psi}}+\bar{p}_{_{\psi}}\right)=\frac{\psi'}{2}\left(\bar{\rho}_{_{\psi}}-\frac{3\bar{p}_{_{\psi}}}{c^{2}}\right)
\end{equation}

Thus  combining  the  two  energy  densities, we  have  from the
above  equations, the  equation  for  the  conservation for the
total  energy  is

\begin{equation}
\bar{\rho'}+3\gamma\bar{H}\bar{\rho}=0
\end{equation}

Here
$\bar{H}=\frac{1}{3}\left(\frac{\bar{a}'}{\bar{a}}+2\frac{\bar{b}'}{\bar{b}}\right)$
is  the  Hubble  parameter  in  the Einstein frame and  $\gamma$
is the  net  barotropic  index  defined  as

\begin{equation}
\gamma\bar{\Omega}=\gamma_{_{f}}\bar{\Omega}_{_{f}}+\gamma_{_{\psi}}\bar{\Omega}_{_{\psi}}
\end{equation}

where

\begin{equation}
\bar{\Omega}=\bar{\Omega}_{_{f}}+\bar{\Omega}_{_{\psi}}=\frac{\bar{\rho}}{3\bar{H}^{^{2}}}
\end{equation}

is  the  dimensionless  density  parameter.\\

From  equations (26)  and  (32), we  have  the  evolution
equation  for  the  density  parameter  as

\begin{equation}
\bar{\Omega}'=\bar{\Omega}(\bar{\Omega}-1)[\gamma\bar{H}_{_{a}}+2(\gamma-1)\bar{H}_{_{b}}]
\end{equation}

where ~~ $\bar{H}_{_{a}}=\frac{\bar{a}'}{\bar{a}}$    and
$\bar{H}_{_{b}}=\frac{\bar{b}'}{\bar{b}}$ .\\

This equation in $\bar{\Omega}$ shows that  $\bar{\Omega}=1$ is a
possible solution  of it and  for stability of  this solution,
we  have

\begin{equation}
\gamma<\frac{2}{3}
\end{equation}

Since  the  adiabatic  indices  do  not  change  due  to
conformal  transformation  so  we  take  $\gamma_{_{f}}=1$ (since
$ p_{_{f}}=0$) and $\gamma_{_{\psi}}=2$. Hence from (33) and (34),
we have

\begin{equation}
\gamma=\frac{\bar{\Omega}_{_{f}}+2\bar{\Omega}_{_{\psi}}}{\bar{\Omega}_{_{f}}+\bar{\Omega}_{_{\psi}}}
\end{equation}

Now  due  to  upper  limit  of  $\gamma$ , we  must  have  the
inequality

\begin{equation}
\bar{\Omega}_{_{f}}<4|\bar{\Omega}_{_{\psi}}|
\end{equation}

according  as  $\gamma$  is  restricted  by  (36).\\

From  the  field equation  (26), the  curvature  parameter
$\bar{\Omega}_{_{k}}=-k c^{2}/~\bar{b}^{2}$ vanishes  for  the
solution $\bar{\Omega}=1$ . So for BD-scalar field it  is possible
to have a stable  solution corresponding  to $\bar{\Omega}=1$ and
hence  the flatness problem  can  be solved .

\section{\normalsize\bf{Concluding  Remarks }}

We  have  performed  an  extensive  analysis  of  solutions  to
Brans-Dicke  theories  with  varying  speed  of  light. For the
power-law  forms  and  exponential  forms  of  the cosmological
scale  factors  and  scalar  field  in  {\it case I} and {\it II}
respectively, we  have  the  velocity  of  light  in the  same
form  and  we  identified  the  cases  where  the quintessence
problem  can  be  solved  for  some  restrictions on  parameters.
In  the  {\it case I}, we  always  have  the accelerated universe.
For figures 1  and  2, $q$  decreases to a  fixed negative value
asymptotically  for  $\omega=-1.8$   and  $\omega=-1.5$ . For
figure 3, $q$ is linear with time  and  always  negative for
$\omega=-1$ . For figure 4, $q$ decreases and  then  there  is  a
singularity and after that $q$ increases. In  figure 5, $q$
increases  to a fixed positive value asymptotically. From figure
6, we conclude that the decelerated parameter  is negative
whenever  $-2<\omega<-2/3$ . In this range $q$  decreases till
$t=-\frac{a_{_{0}}(3\omega+2)}{6a_{_{1}}(\omega+1)}$ and
then  increases  to  a  fixed  positive  value asymptotically.\\

Along  with  providing  a  non-decelerating  solution, it  can
solve  the  flatness  problem  also. In  fact, it  has  been
shown  that  $\bar{\Omega}=1$   could  be  a  stable  solution  in
this model.\\

{\bf Acknowledgement:}\\

The  authors  are  thankful  to  the  Relativity  and  Cosmology
Research  Centre, Department  of  Physics, Jadavpur  University
for  helpful  discussion. One  of  the authors (U.D)  is
thankful  to  CSIR  (Govt. of  India)  for  awarding  a  Junior
Research Fellowship.\\

{\bf References:}\\
\\
$[1]$  N. A. Bachall, J. P. Ostriker, S. Perlmutter, P.
J. Steinhardt, {\it Science} {\bf 284} (1999) 1481.\\
$[2]$  Perlmutter S et al 1998 {\it Nature} (London) {\bf 391}
51; 1999 {\it Astrophys  J} {\bf 517} 565.\\
$[3]$  Riess A G  et al 1998 {\it Astrophys  J} {\bf 116} 1009;
Garnavich P M  et al  1998 {\it Astrophys  J} {\bf 509} 74.\\
$[4]$  R. R. Caldwell, R. Dave  and  P. J. Steinhardt, {\it Phys.
Rev. Lett.} {\bf 80}, 1582 (1998) [{\it astro-ph} / 9708069].\\
$[5]$  V. Sahni, A. A. Strarobinsky, {\it Int. J. Mod. Phys. D}
{\bf 9} 373 (2000)  [{\it astro-ph}/9904398].\\
$[6]$  A. A. Strarobinsky, {\it JEPT Lett.} {\bf 68} (1998) 757;
T. D. Saini, S. Roychaudhury, V. Sahni, A. A. Strarobinsky, {\it
Phys. Rev. Lett.} {\bf 85} (2000) 1162.\\
$[7]$  Faraoni V  2000 {\it Phys. Rev. D} {\bf 62}  023504.\\
$[8]$  Bucher M and Spergel  D  1999 {\it Phys. Rev. D} {\bf 60}
043505.\\
$[9]$  Battye R A, Bucher M  and Spergel D "Domain  wall dominated
universe" ,{\it astro-ph}/990847.\\
$[10]$  L. Wang, R. Caldwell, J. P. Ostriker  and  P. J.
Steinhardt, {\it Astrophys. J.} {\bf 530} 17 (2000).\\
$[11]$  I. Waga and P. M. R. Miceli, {\it Phys. Rev. D} {\bf 59}
103507 (1999).\\
$[12]$  Chimento L  P, Jakubi  A  S  and Pavo'n D 2000, {\it
Phys. Rev. D} {\bf 62} 063508 (Preprint  {\it astro-ph}/005070).
$[13]$  Banerjee  N  and  Pavo'n  D  2001  {\it Phys. Rev. D}
{\bf 63} 043504.\\
$[14]$  A. Albrecht  and  J. Magueijo, {\it Phys. Rev. D} {\bf 59}
043516 (1999). \\
$[15]$  J. D. Barrow, {\it Phys. Rev. D} {\bf 59} 043515 (1999).\\
$[16]$  J. D. Barrow  and  J. Magueijo, {\it Phys. Lett. B} {\bf
443} 104 (1998); {\it Phys. Lett. B} {bf 447} 246 (1999); {\it
Class. Quantum  Grav.} {\bf 16} 1435  (1999). \\
$[17]$  J. D. Barrow and J. Magueijo, {\it Astrophys. J. Lett.}
{\bf 532} L87 (2000); J. Magueijo, {\it Phys. Rev. D} {\bf 62}
103521 (2000). \\
$[18]$  J. Moffat, {\it Int. J. Mod. Phys. D} {\bf 2} 351 (1993);
{\it Fund. Phys.} {\bf 23}  411  (1993). \\
$[19]$  M. A. Clayton and J. W. Moffat, {\it Phys. Lett. B} {\bf
460} 263 (1999); {\it Phys. Lett. B} {\bf 477} 269 (2000). \\
$[20]$  E. Kiritsis, {\it J. High  Energy  Phys.} {\bf 10} 010
(1999). \\
$[21]$  J. Magueijo, {\it Phys. Rev. D} {\bf 63} 043502 (2001).\\
$[22]$  Thorne  K S 1967  {\it Astrophys. J.} {\bf 148}  51. \\
$[23]$  Faraoni V, Gunzig  E and  Nardone  P  1999, {\it Fundam.
Cosm. Phys.} {\bf 20} 121.\\

\end{document}